\documentclass[twocolumn,showpacs,preprintnumbers,amsmath,amssymb]{revtex4}

\usepackage{graphicx}
\usepackage{dcolumn}
\usepackage{bm}

\begin{document}


\title{Decay of Langmuir wave in dense plasmas and warm dense matter}

\author{S. Son}
\affiliation{18 Caleb Lane, Princeton, NJ, 08540}
\email{seunghyeonson@gmail.com}
\author{S. Ku}
\email{sku@cims.nyu.edu}
\affiliation{Courant Institute of Mathematical Sciences, New York University, New York, NY 10012}
\author{Sung Joon Moon}
\affiliation{PACM, Princeton University, Princeton, NJ 08544}
\date{\today}

\begin{abstract}
  The decays of the Langmuir waves in dense plasmas are computed using  the dielectric function theory  recently proposed~\cite{sonpre}. 
   Four cases are considered: a classical plasma, a Maxwellian plasma, a degenerate quantum plasma, and a partially degenerate plasma. The results obtained suggest considerable deviations from the conventional Landau damping theory. 
 Its implication  on the x-ray Raman compression in dense plasmas or warm dense matter is discussed.   
\end{abstract}



\maketitle

\section{Introduction}

The decay of the Langmuir wave is one of the most important physics in plasmas related  to many applications \cite{surface, photo, nano, wake, Fisch, Fisch2, sonpre}.  
While the Landau damping theory is accurate to predict the plasmon decay in ideal classical plasmas, it is inadequate in dense plasmas and warm dense matter. For an example, in metals,  the experimental measurement shows that the decay of the long-wave length plasmon is  higher than that predicted by the Landau damping theory \cite{Gibbons}.  
 Previously, a new theory~\cite{sonpre2}, which is a natural extension from the one in the condensed matter~\cite{Sturm, Sturm2, Mermin},  has been proposed to compute the plasmon decay in dense plasmas.  

In this paper, we provide detailed calculations of  this theory \cite{sonpre2}. 
We consider four cases: a classical plasma, a Maxwellian plasma, a degenerate quantum plasma, and a  partially degenerate plasma. The decays in the limit, where the wave vector goes to zero, are finite in all cases.  The decay rate is lower in finite-temperature plasmas than degenerate plasmas.  We then discuss the implication of the result  on the x-ray Raman compression where dense plasmas or warm dense matter are used as a compressing media.  

This paper is organized as follows. In section II, a new theory \cite{sonpre2} is introduced to predict the plasmon decay in dense plasmas and warm dense matter. In section III, we compute the decay rate implied by the theory \cite{sonpre2} in the limit when $\hbar=0$, where $\hbar$ is the Frank constant. In section IV,  we compute the plasmon decay for a Maxwellian plasma when $\hbar \neq 0$.  In section V, the plasmon decay is computed for the quantum electron plasma ($\hbar \neq 0$) when $T_e=0$.  In section VI, the plasmon decay is computed when $T_e\neq 0$ and $\hbar\neq 0$.  In section VII, the summary is given, and we discuss the implication of the result obtained on the x-ray Raman compression in dense plasmas.

\section{Dielectric Function  Theory}

Previously, it is shown that the dielectric function in dense plasmas under a wave with wavevector $\mathbf{k}$ and angular frequency $\omega$ has additional term~\cite{sonpre2}: 

\begin{equation}
  \epsilon(\mathbf{k},\omega)  = 1 + \frac{4 \pi e^2 }{ k^2 } \alpha_{\mathrm{rpa}}(\mathbf{k},\omega) + 
\frac{4 \pi e^2 }{ k^2 } \alpha_{\mathrm{dense}}(\mathbf{k}, \omega) \mathrm{,}
\end{equation}
where $ \alpha_{\mathrm{dense}}$ is the special term that is significant in dense plasmas. 
The $\alpha_{\mathrm{rpa}}$ is the well-known Lindhard susceptibility \cite{Lindhard} 
\begin{equation}
\alpha_{\mathrm{rpa}} =   \int \frac{d^3 \mathbf{q}}{ (2\pi)^3 } \frac{f(|\mathbf{k}+\mathbf{q}|) - f(\mathbf{q}) }{ \hbar \omega - E(\mathbf{k}+\mathbf{q}) + E(\mathbf{k}) } \mathrm{,} \label{eq:lind}
\end{equation}
where $f$ is the occupation number and  $E(q) = \hbar^2 q^2 / 2m_e$.   The $\alpha_{\mathrm{dense}}$ is computed \cite{sonpre2} as 
\begin{eqnarray}
\alpha_{\mathrm{dense}}(\mathbf{k}, \omega) = \int \frac{d^3 \mathbf{s}}{(2\pi)^3} n_i(s)  \int \frac{d^3 \mathbf{q}}{ (2\pi)^3} \frac{U^2(s)}{A^2(\mathbf{q},\mathbf{k},\mathbf{s})} \nonumber \\
\times 
 \frac{ f(|\mathbf{q}+\mathbf{k}+\mathbf{s}|) - f(\mathbf{q})}{ \hbar \omega  - E(|\mathbf{q}+\mathbf{k}+\mathbf{s}|) + E(\mathbf{q})} \mathrm{,} \nonumber \\ \label{eq:eigen3}
\end{eqnarray}
where  $ U(s) = 4 \pi Z_i e^2 / (s^2 + k_{s}^2) $ is the screened ion-electron potential with screening length  $1/ k_{s}$,  $n_i(s)$ is the ion-ion structure factor $  n_i(s)  =  \langle \sum_{i,j} \exp(i\mathbf{s}\cdot (\mathbf{X}_i-\mathbf{X}_j))/V$,  $ A(\mathbf{q},\mathbf{k},\mathbf{s})$ is defined as 
\begin{eqnarray}
A^{-1}(\mathbf{q},\mathbf{k},\mathbf{s}) = \frac{1}{E(\mathbf{q}) - E(|\mathbf{q}+\mathbf{s}|)}\nonumber \\
 - \frac{1}{E(|\mathbf{q}+\mathbf{k}|) - E(|\mathbf{q}+\mathbf{k}+\mathbf{s}|)} \mathrm{.}  \nonumber \\ \label{eq:a}
\end{eqnarray}
The imaginary part of $\alpha_{\mathrm{dense}}$ is given as

\begin{eqnarray}
\mathrm{Im}\left[ \alpha_{\mathrm{dense}}\right] = 
\int \frac{d^3 \mathbf{s}}{(2\pi)^3} n_i(\mathbf{s})  \int \frac{d^3 \mathbf{q}}{ (2\pi)^3}\frac{U^2(s)}{A^2(\mathbf{q},\mathbf{k},\mathbf{s})}  \nonumber \\
\times  \pi \delta ( \hbar \omega  - E(|\mathbf{q}+\mathbf{k}+\mathbf{s}|) + E(\mathbf{q})) \nonumber \\ 
\times \left(f(|\mathbf{q}+\mathbf{k}+\mathbf{s}|) - f(\mathbf{q})\right)\mathrm{.} \nonumber \\
\label{eq:eigen4}
\end{eqnarray}

\section{Classical Plasmas: $\hbar = 0$} 
\begin{figure}
\scalebox{0.6}{
\includegraphics{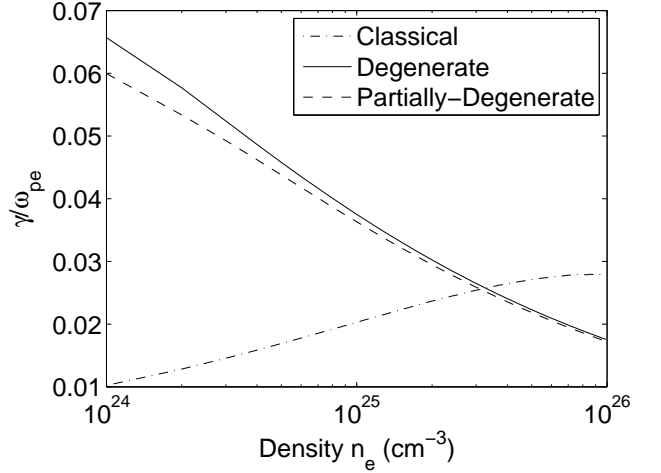}}
\caption{\label{fig:1} Damping rates of classical 400 eV hydrogen plasma (Classical), degenerate plasma (Degenerate), and partially degenerate plasma (Partially-degenerate),  where   $ 10^{24}/ \mathrm{cc} < n_e <  10^{26} /\mathrm{cc}$ and $k \gg k_F$. The electron temperature of the partially degenerate plasma is $0.5 E_f(n_e) $.  The damping rate $\gamma$ is normalized by $\omega_{\mathrm{pe}}$. 
}
\end{figure}

In the limit $\hbar =0$,  the $A$ in Eq.(\ref{eq:a}) is simplified as 

\begin{equation} 
A^{-1}(\mathbf{k},\mathbf{s}, q_2) = \frac{ \mathbf{k} \cdot \mathbf{s}}{m_e\omega^2}  \kappa^{-} \kappa^{+} \label{eq:sp1} \mathrm{,} 
\end{equation}
where 
\begin{equation}
\kappa^{\pm}(k,s, \cos(\theta))   = \left( 1 - \frac{k^2 + sk \cos(\theta)}{|\mathbf{s}+\mathbf{k}|^2} \right)^{-1} \nonumber  \mathrm{,}
\end{equation}
and $\cos(\theta)$ is the angle between $\mathbf{k}$ and $\mathbf{s}$. 
From Eqs. (\ref{eq:eigen3}) and (\ref{eq:sp1}),  the Eq.(\ref{eq:eigen4}) can be simplified as

\begin{eqnarray} 
 \frac{4 \pi e^2}{k^2} \mathrm{Im} \left[ \alpha_{\mathrm{dense}} \right] 
=  \int \frac{n_i(\mathbf{s}) d^3 \mathbf{s}}{ (2\pi)^3} 
 \left( \kappa^+(\mathbf{k},\mathbf{s}) \kappa^-(\mathbf{k},\mathbf{s}) \right)^2 \nonumber \\
\times  \left( \frac{\mathbf{k} \cdot \mathbf{s} }{ m_e \omega^2}\right)^2\left(\frac{4 \pi Z_i e^2}{s^2 + k_{\mathrm{de}}^2} \right)^2 
 \pi \left(\frac{\omega}{k} \right)^2 \frac{\partial f(\omega/|\mathbf{k}+\mathbf{s}|)}{ \partial v} \mathrm{,} \nonumber \\ \nonumber 
\end{eqnarray}
where $k_{\mathrm{de}} = 4 \pi n_e e^2 / T_e $ is the Debye screening length. By using the spherical coordinate,   it can be re-casted as 

\begin{equation} 
 \frac{4 \pi e^2}{k^2} \mathrm{Im} \left[ \alpha_{\mathrm{dense}} \right] 
= \frac{ \sqrt{2\pi}}{ 8 \pi^2} \left( \frac{Z_i }{ n_e\lambda_{\mathrm{de}}^3 } \right)  \log(\Lambda) \mathrm{,}  \label{eq:cla}
\end{equation} 
where the logarithmic factor, $\log(\Lambda) $, is given as 
\begin{eqnarray}
\log(\Lambda) = \int_0^{\infty} \frac{  d s}{|\mathbf{k}+\mathbf{s}|} \int_{-1}^{+1} u^2 du  \frac{n_i(\mathbf{s}) }{\bar{n}_i} 
 \left(\frac{s^2}{s^2 + k_{\mathrm{de}}^2}\right)^2 
  \nonumber \\ \nonumber \\ \nonumber  \times  \left(  \kappa^{-}\kappa^{+}\right)^2
 \exp(-\frac{\omega^2}{ 2  |\mathbf{k}+\mathbf{s}|^2 v_{\mathrm{te}}^2} ) 
\mathrm{,} \ \nonumber 
\end{eqnarray}
and $ u = \cos(\theta)$, $v_{\mathrm{te}} = \sqrt{ T_e/m_e}$ and $\bar{n}_i$ is the ion average density. In particular,  in the limit $k=0$,  $ A^{\pm} = 1 $, the logarithm factor is given as  
\begin{eqnarray}
\log(\Lambda) = \int_0^{\infty} \frac{d s}{s}  \frac{2}{3}  \frac{n_i(s) }{\bar{n}_i} 
 \left(\frac{s^2}{s^2 + k_{\mathrm{de}}^2}\right)^2
 \exp(-\frac{\omega^2}{ 2  s^2  v_{\mathrm{te}}^2} ) \mathrm{.}  \nonumber 
\end{eqnarray}
This integration is divergent in the $s$-integration due to the lack of the high cutoff.  We introduce the high cutoff as $ s_{\mathrm{high}} = 1/b_c$ where $b_c = e^2 / T_e $ is the closest approaches.   Since the decay rate is proportional to $n_i$ and $T_e^{-3/2}$, its rough interpretation should be the ion-electron collision.   
 In Fig.~\ref{fig:1}, we show the decay rate, in the limit $k \cong 0$, normalized by $\omega_\mathrm{pe}$ as a function of the electron density for the long wave length plasmon, where $\omega_\mathrm{pe}=(4\pi n_e e^2/m_e)^{1/2}$.

\section{Maxwellian Plasma: $\hbar \neq 0$}
\begin{figure}
\scalebox{0.6}{
\includegraphics{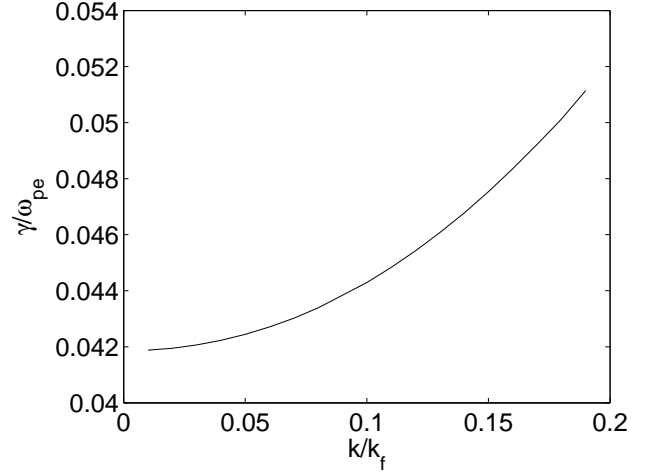}}
\caption{\label{fig:2} Damping rate of 36 eV hydrogen plasma as a function of $q$, where   $n_e =  10^{24} /\mathrm{cc}$. The x-axis is the wave vector $k/k_F$ and the y-axis is damping rate $\gamma/\omega_{\mathrm{pe}}$. 
}
\end{figure}

In this section, we consider the case when the electron distribution, $f$, is given as a Maxwellian and ignore the Fermi degeneracy. In this case, the susceptibility of the Lindhard function in Eq.(\ref{eq:lind}) is given as   

\begin{equation}
\alpha_{\mathrm{rpa}} =  \int \left[\frac{f_m(\mathbf{v}+\frac{\hbar \mathbf{k}}{2 m_e}) - f_m(\mathbf{v}-\frac{\hbar \mathbf{k}}{2 m_e})}{\hbar (\omega -\mathbf{k} \cdot \mathbf{v})} \right] d^3 \mathbf{v} \mathrm{,} 
\end{equation}
where $f_m = 1/(2 \pi v_{\mathrm{te}}^2)^{3/2} \exp( -(v^2/ 2 v_{\mathrm{te}}))$, and $ \mathbf{v} = \hbar \mathbf{q} / m_e$ is the electron velocity.   
In case when $k$ is not zero, the $A$ in Eq.(\ref{eq:eigen4}) is dependent on $\mathbf{q}$. The rather complicated expression for $A$ is given as 

\begin{eqnarray}
A^{-1}(\mathbf{k}, \mathbf{s}, \mathbf{q} ) &=& 
\frac{\hbar^2 \mathbf{k} \cdot \mathbf{s}}{m_e(\hbar \omega)^2} 
\left( 1 - \frac{\hbar \mathbf{k} \cdot (\mathbf{q} + \mathbf{s}/2)}{ m\omega} \right)^{-1} \nonumber \\
&\times& \left( 1 - \frac{\hbar \mathbf{k} \cdot (\mathbf{q} - \mathbf{s}/2)}{m \omega} \right)^{-1} \nonumber 
\end{eqnarray} 
$A$ cannot be taken out of the $\mathbf{q}$-integration in Eq.~(\ref{eq:eigen3}) due to its dependency on $\mathbf{q}$.  For a given $\mathbf{s}$ and $\mathbf{k}$, we do the $\mathbf{q}$-integration first and then $\mathbf{s}$-integration later.   For simplicity, assume $\mathbf{k} = (k_x, 0, 0)$ and $\mathbf{s} = (s\cos(\theta), s \sin(\theta), 0)$. Let us represent the $\mathbf{q} = q_1 \hat{x}_1
+ q_2 \hat{x}_2 + q_3 \hat{x}_3$, where $\hat{x}_1 = (k+ s \cos(\theta), s\sin(\theta), 0)/|k+s|$, $ \hat{x}_1 =  (- s\sin(\theta), 1 + \cos(\theta), 0)/|k+s|$,  and $ \hat{x}_3 = (0,0, 1)$. In this coordinate system,  $ A $ is only dependent on $k$, $s$, $\theta$, and $q_2$: 

\begin{equation} 
A^{-1}(\mathbf{k},\mathbf{s}, q_2) = \frac{ \mathbf{k} \cdot \mathbf{s}}{m_e \omega^2}  \kappa^{-} \kappa^{+} \label{eq:sp}  \mathrm{,} 
\end{equation}
where 
\begin{equation}
\frac{1}{\kappa^{\pm}}  = 1 - \frac{k^2 + sk \cos(\theta)}{|\mathbf{s}+\mathbf{k}|^2}
+ \frac{\hbar}{m_e \omega}( 
\frac{sk\sin(\theta) q_2}{|\mathbf{s} + \mathbf{k}|} \pm  \frac{sk \cos(\theta)}{2})  \nonumber  \mathrm{.}
\end{equation}
 The real part of $\alpha$ in Eq.(\ref{eq:eigen4}) is still very complicated, but the imaginary part given in Eq.(\ref{eq:eigen4}) could be simplified due to the fact that the delta function will eliminate  $q_1$ integration and the integrand of $q_3$ can be done easily:

\begin{eqnarray}
  \int \frac{d^3 \mathbf{q}}{ (2\pi)^3}\frac{U^2(s)}{A^2(\mathbf{q},\mathbf{k},\mathbf{s})} \mathrm{Im} \left[ \frac{f(\mathbf{q}+\mathbf{k}+\mathbf{s}) - f(\mathbf{q})}{  \hbar \omega  - E(\mathbf{q}+\mathbf{k}+\mathbf{s}) + E(\mathbf{q})}\right] \nonumber \\  \nonumber \\ \nonumber \\
= \frac{( \mathbf{k} \cdot \mathbf{s})^2 U^2(s)}{m_e^2\omega^4}  \int_{-\infty}^{\infty} dq_2  \left(f_m(Q^+, q_2,\omega) - 
f_m(Q^-, q_2, \omega) \right) \nonumber \\ \times 
  \frac{\pi m_e}{\hbar^2 |\mathbf{s}+\mathbf{k}|}( \kappa^{+} \kappa^{-1})^2  \nonumber 
 \mathrm{,} \nonumber \\  \label{eq:ccc}
\end{eqnarray}
where $Q^{\pm}(|k+s|) =  m_e\omega/\hbar|s+k| \pm |k+s|/2$. 
Note that $f_m$ is the Maxwellian distribution with $q_3$ having been integrated out: 
\begin{equation}
f_m(q_1, q_2, \omega)  = \frac{1}{ 2 \pi k_{\mathrm{te}}^2}
\exp( -\frac{ q_2^2 + q_1^2 }{ 2 q^2_{\mathrm{te}}})\mathrm{,} \nonumber 
\end{equation}
where $ q_{\mathrm{te}} =  \sqrt{m_eT_e/\hbar^2} $. 
 The right-hand side of Eq.~(\ref{eq:ccc})  is only a function of $k$, $s$, $\theta$. We could do the $\mathbf{s}$-integration in a spherical coordinate to have the final expression as

\begin{eqnarray}
 \mathrm{Im} \left[\alpha(\mathbf{k}, \omega)\right] = \int \frac{2\pi s^2 d\mu ds}{(2\pi)^3 } n_i(s) \frac{( ks \mu)^2}{m_e^2\omega^4} 
\left( \frac{4 \pi Z_i e^2 }{s^2  + \lambda_{\mathrm{de}}^{-2} } \right)^2 
 \nonumber  \\
\times \int_{-\infty}^{\infty} dq_2  \left(f_m(Q^+, q_2\omega) - f_m(Q^-, q_2, \omega) \right) \nonumber \\ 
\times \frac{\pi m_e}{\hbar^2 |\mathbf{s}+\mathbf{k}|} (\kappa^{+} \kappa^{-1})^2 \mathrm{.~~~~~}  \label{eq:qcla}
\end{eqnarray}
In Fig.~\ref{fig:2}, we plot the decay rate as a function of $k$ when $n_e = 10^{25} / \mathrm{cc} $.

\section{Degenerate Case:  $\hbar \neq 0 $ and $T_e = 0$} 

Consider a dense plasma, where electrons are completely degenerate.   The Fermi-energy is given as
$ E_F =  36.4 \times (n/n_{24})^{2/3} \mathrm{eV}$ where  $n_{24} = 10^{24} / \mathrm{cc}$, and as long as $T_e \ll E_F$, the computation in this section will be valid.     
To begin with, the susceptibility $\alpha_{\mathrm{rpa}} $ of the free electron plasma has been computed by Lindhard \cite{Lindhard} and is given as 
\begin{equation}
\alpha_{\mathrm{rpa}}(\mathbf{k},\omega) = \frac{3 n_e}{m_e v_F^2} h(z, u) \mathrm{,} \nonumber 
\end{equation}    
where $v_F = \sqrt{2E_F/m_e}$ is the Fermi velocity, $ z = k / 2 k_F$, $u = \omega /k v_F$, and $h= h_r + i h_i$. The real part of $h$ is given as 
\begin{eqnarray}
h_r = \frac{1}{2} + \frac{1}{8z}\left( 1- (z-u)^2\right) 
\log \left( \frac{|z-u+1|}{|z-u+1|} \right) \nonumber  \\
+ \frac{1}{8z}\left( 1- (z+u)^2\right) 
\log \left( \frac{|z+u+1|}{|z+u+1|} \right)\nonumber \\ \label{eq:re} \nonumber 
\end{eqnarray}
The imaginary part of $h$ is given as 
\begin{eqnarray}
h_i  =&  \frac{\pi}{2} u \mathrm{,}~             &\mathrm{if}  |z+u| < 1   \mathrm{,} \nonumber \\
  &  \frac{\pi}{8z} (1-(z-u)^2) \mathrm{,}~  &\mathrm{if} |z-u| < 1 <  |z+u|  \mathrm{,}\nonumber \\ 
    &  0   \mathrm{,}~                         &\mathrm{if}  |z-u| > 1 \mathrm{.} \nonumber 
 \end{eqnarray}
In the limit where $k =0 $,  while computing the imaginary part of $\alpha_{\mathrm{dense}}$,  
$A$ can be approximated as 
$  A^{-1} = (\mathbf{k} \cdot \mathbf{s})/m_e \omega^2$, due to the delta function in Eq.(\ref{eq:eigen4}). 
$A$ becomes independent to $\mathbf{q}$ and it can be taken out from the $\mathbf{q}$-integration in Eq.(\ref{eq:eigen4}). 
Using Eq.~(\ref{eq:lind}), Eq.(\ref{eq:eigen4}) can be simplified to 
\begin{eqnarray}
 \mathrm{Im} \left[\alpha_{\mathrm{dense}}(\mathbf{k},\omega)\right] =   \int \frac{d^3 \mathbf{s}}{ (2\pi)^3 } \frac{(\mathbf{k} \cdot \mathbf{s})^2 (4 \pi Z_ie^2)^2 }{m_e^2 \omega^4 (s^2 + k_{tf}^2)^2 } \nonumber \\ \nonumber \\
\times n_i(s) \mathrm{Im} \left[ \alpha_{\mathrm{rpa}}(\mathbf{k} + \mathbf{s}, \omega) \right] \mathrm{.}\nonumber \\ \label{eq:zero2} \nonumber 
\end{eqnarray}
The above equation can be further simplified to 
\begin{eqnarray}
 \mathrm{Im} \left[\alpha_{\mathrm{dense}}(\mathbf{k}, \omega)\right]
= \frac{4 E_F^2 Z_i (k_F e^2)^2 }{(\hbar\omega)^4 }\frac{(4\pi)^2}{ 3 \pi^2 } \int \frac{d^3 \mathbf{s}}{ k_F^3(2\pi)^3 }   \nonumber \\ \nonumber \\
\times \frac{(\mathbf{k}\cdot \mathbf{s})^2}{ k^2 (s^2 + k_{tf}^2)^2 } \mathrm{Im} \left[\alpha_{\mathrm{rpa}} (|k+s|, \omega) \right] \mathrm{,}  \label{eq:kzero}
\end{eqnarray}
where we assumed $n_i(\mathbf{s}) = n_i$. 
For $U(s)$, we will use the  screening potential $ U(s) = 4 \pi Z_i e^2 / (s^2 + k_{tf}^2) $, where $k_{tf} = \sqrt{3 \omega_{\mathrm{pe}}^2/v_F^2} $ is the Thomas-Fermi screening length. In Fig.~\ref{fig:1}, we plot  $\gamma /\omega_{\mathrm{pe}}$ as  a function of electron density $n_e$ using the above equation (Degenerate).

\begin{figure}
\scalebox{0.6}{
\includegraphics{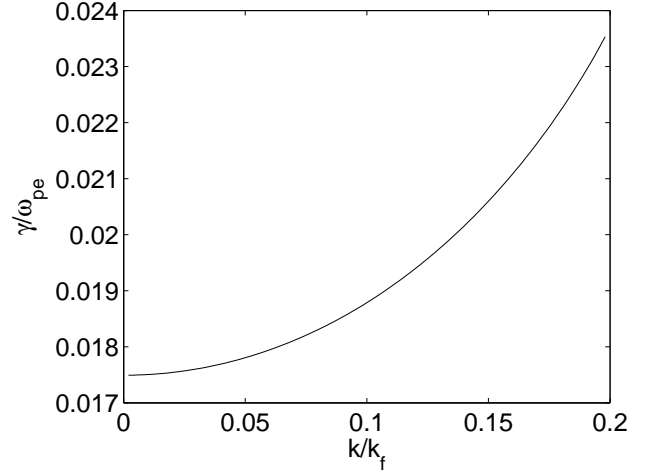}}
\caption{\label{fig:4} Damping rates of a degenerate hydrogen plasma as a function of the wave vector when $n_e = 10^{26} /\mathrm{cc}$. The x-axis is wave vector $k/k_F$ and the y-axis is the damping rate $\gamma/\omega_{\mathrm{pe}}$. 
}
\end{figure}

In the case of $k \neq 0 $, the above simplification is not possible.  Using Eq.(\ref{eq:sp}), we need the $q_1$, $q_2$ and $q_3$ integrations. The real part of the susceptibility in Eq.~(\ref{eq:eigen3}) is complicated but the imaginary part given in Eq.~(\ref{eq:eigen4}) is simpler due to the delta function.  The $q_1$ integration is eliminated by the delta function and, the $q_3$-integration can be done since the integrand is independent of $q_3$. 
 We will not present the detailed steps,  but, after tedious manipulation, the $q$ integration can be simplified to an 1-dimensional integration given by

\begin{eqnarray}
  \int \frac{d^3 \mathbf{q}}{ (2\pi)^3}\frac{U^2(s)}{A^2(\mathbf{q},\mathbf{k},\mathbf{s})} \mathrm{Im} \left[ \frac{f(\mathbf{q}+\mathbf{k}+\mathbf{s}) - f(\mathbf{q})}{  \hbar \omega  - E(\mathbf{q}+\mathbf{k}+\mathbf{s}) + E(\mathbf{q})}\right] \nonumber \\  \nonumber \\ \nonumber \\
= \frac{(\hbar^2 \mathbf{k} \cdot \mathbf{s})^2 U^2(s)}{m_e^2(\hbar \omega)^4}  \int_{-k_F\sqrt{1-(z-u)^2}}^{k_F\sqrt{1-(z-u)^2}} dq_2 
  \frac{m_e}{4\pi^2 \hbar^2 |\mathbf{s}+\mathbf{k}|}  \nonumber 
 \\ \nonumber \\ \times \kappa^{+} \kappa^{-1} \left(R^-(k_F, z, u, q_2)-R(k_f, z,u, q_2)\right) \mathrm{,} \nonumber \\  \label{eq:comp} 
\end{eqnarray}
where $ R^{\pm}(k_F,z,u,q_2)  =  \sqrt{ k_F^2(1 - (z\pm u)^2 ) - q_2^2 }$,  $z = |k+s|/ 2k_F$ and  $u = \omega / kv_F$. The right-hand side of the above equation is only a function of $k$, $s$, $\theta$. We could do the $\mathbf{s}$-integration in spherical coordinate to have the final expression as

\begin{eqnarray}
 \mathrm{Im} \left(\alpha(\mathbf{k}, \omega)\right) = \int \frac{2\pi s^2 d\mu ds}{(2\pi)^3 } n_i(s) \frac{(\hbar^2 ks \mu)^2 U^2(s)}{m_e^2(\hbar \omega)^4} \nonumber \\ \nonumber \\
\times  \int_{-\sqrt{1-(z-u)^2}}^{\sqrt{1-(z-u)^2}} dq_2   \frac{m_e}{4\pi^2 \hbar^2 |s+k|} \nonumber \kappa^{+} \kappa^{-1} \nonumber  \\  \nonumber \\ \times 
   \left(R^-(k_F, z, u, q_2)-R(k_f, z,u, q_2)\right)  \mathrm{,}\nonumber \\  \label{eq:comp2}  
\end{eqnarray}
 In Fig.~\ref{fig:4}, we plot  $\gamma/ \omega_{\mathrm{pe}} $ in Eq.(\ref{eq:comp2})  as a function of $k$ for a hydrogen plasma with the electron density of $n_e= 10^{26} / \mathrm{cc}$.

\section{Partially Degenerate Case:  $\hbar \neq 0 $ and $T_e \neq 0$}
Due to the partial degeneracy and quantum diffraction,  this case is very hard to treat.  In Eq.(\ref{eq:eigen4}),  the $q_3$-integration cannot be simplified due to the partial degeneracy.   However, there is some simplification achievable due to the Dharma-Wardana's technique \cite{wardana}.   For partially degenerate electrons,  the occupation number is given as

\begin{equation} 
f(T_e, n_e, E) = \left[ \exp\left( \frac{E-\mu(n_e, T_e)}{T_e} \right)  +1 \right]^{-1}\label{eq:mermin} \mathrm{.}
\end{equation}
where $E$ is the electron kinetic energy, $\mu$ is the chemical energy. 
Dharm-Wardana showed that the $f$ in Eq.(\ref{eq:mermin}) could be expanded as follows. 

\begin{eqnarray}
f(T_e, n_e, E) &=& \int_{0}^{\infty} \left[ 2 T_e \left( 1+ \cosh\left(\frac{\mu - S}{T_e}\right)\right)\right]^{-1} \nonumber \\  
&~&\times g(0, n_{m}(S), E ) dS \mathrm{,~~~~} \label{eq:mermin2} 
\end{eqnarray}
where  $n_{m}(S) = (1/3\pi^2) (2m_eS/\hbar^2)^{3/2}$ is the electron density whose Fermi energy is $S$, and $  g(0, n_{m}(S), E )$ is the occupation number for the zero temperature plasma with the electron density of $ n_{m}(S) $.  
The Eq.~(\ref{eq:mermin2}) makes it possible to express the dielectric function of the electrons  as the sum of the dielectric functions of the zero temperature electrons. 
For an example, the susceptibility of the Lindhard dielectric function for electrons with a  non-zero temperature could be expressed as an integral of the susceptibility of the zero-temperature electrons: 

\begin{eqnarray} 
\alpha_{\mathrm{rpa} }(\mathbf{k},\omega, T_e, n_e) = 
  \int_{0}^{\infty}  \alpha_{\mathrm{rpa} }(\mathbf{k},\omega, 0, n_m(S)) \nonumber \\ \nonumber  \\ \nonumber 
\times   \frac{1}{ 2 T_e ( 1+ \cosh(
\frac{\mu(n_e, T_e) - S}{T_e}))} dS \mathrm{.} \\ \nonumber 
\nonumber
\end{eqnarray}
We can apply the same technique. First, define, 

\begin{eqnarray}
 B(\mathbf{k}, \omega, 0, n_e)  =   \int \frac{d^3 \mathbf{q}}{ (2\pi)^3}\frac{U^2(s)}{A^2(\mathbf{q},\mathbf{k},\mathbf{s})}\nonumber \\ 
\nonumber \\ \nonumber \times  \mathrm{Im}
 \left[ \frac{f(\mathbf{q}+\mathbf{k}+\mathbf{s}) - f(q)}{  \hbar \omega  - E(\mathbf{q}+\mathbf{k}+\mathbf{s}) + E(\mathbf{q})}\right] \mathrm{.}
\end{eqnarray}
The $B(\mathbf{k}, \omega, 0, T_e)$ is given in Eq.~(\ref{eq:comp}). Then, using Eqs. (\ref{eq:eigen4}), (\ref{eq:comp}) and (\ref{eq:mermin2}),  the $\mathrm{Im} (\alpha_{\mathrm{dense}})$ is given as 

\begin{eqnarray}
 \mathrm{Im} \left(\alpha_{\mathrm{dense}}\right)
 = \int_{0}^{\infty} \frac{1}{ 2 T_e ( 1+ \cosh(
\frac{\mu(n_e, T_e) - S}{T_e}))}\nonumber \\ \nonumber   \times 
\left[ \int \frac{d^3 \mathbf{s}}{(2\pi)^3} n_i(s)
    B(\mathbf{|k+s|}, \omega, 0, n_m(S))\right] dS  \nonumber \\ \label{eq:partial}
\end{eqnarray}

In particular,  in the limit of $k=0 $ or with the approximation that $\kappa^{\pm}=1$ in Eq.~(\ref{eq:sp}), the damping rate could be obtained from  Eq.(\ref{eq:kzero}) by replacing $\epsilon_{\mathrm{rpa}}( |k + s|, \omega, 0, n_e) -1$: 

\begin{eqnarray}
\epsilon_{\mathrm{rpa}}( T_e, n_e) - 1  =
  \int_{0}^{\infty}  (\epsilon_{\mathrm{rpa} }(0, n_m(S)) -1) \nonumber \\ \nonumber  \\ \nonumber 
\times   \frac{1}{ 2 T_e ( 1+ \cosh(
\frac{\mu(n_e, T_e) - S}{T_e}))} dS \mathrm{.} \\ \label{eq:comp3}
\end{eqnarray} 
In Fig.~\ref{fig:1},  the decay rate of the long wave-length plasmon, using Eq.(\ref{eq:partial}), is plotted as a function of $n_e$ for a hydrogen plasma.

\section{Conclusion}

The Landau damping theory is inadequate in dense plasmas; The decay rate observed from the electron stopping experiment in metals is  finite. 
Previously,  a new theory  \cite{sonpre2}  has been proposed to predict more accurately the plasmon decay in dense plasmas. In this paper, 
we provide the detailed calculations of the theory \cite{sonpre2} for various regime of dense plasmas.

First, we consider classical plasmas where $\hbar = 0 $.  The damping rate resembles the electron collision rate, which is proportional to $n_i$ and $T_e^{-3/2}$.  According to the theory,  the $\gamma/\omega_{\mathrm{pe}}$ is, for a fixed temperature,  an increasing function of $n_e$  as shown in Fig.~(\ref{fig:1}).  
In the integration, there is a divergence in the high $k$, which could be avoided by introducing the cutoff given by the closest approach.  The theory is be valid if the de Broglie wave length is smaller than the closest approach.   
Second, we consider a Maxwellian plasma.  Contrary to the classical plasma, the high cutoff is not needed since it is provided by the thermal De Broglie wave length. This theory is useful for hot dense plamsas, where the quantum diffraction is not negligible, but the quantum degeneracy is.  
Third, we consider a completely degenerate plasma.  Due to the degeneracy, $\gamma/\omega$ is a decreasing function of $n_e$ as shown in Fig.~(\ref{fig:1}), which  is an opposite case with the one in  classical plasmas (as the electron density gets higher, the classical theory breaks down since it fails to take into account the degeneracy). 
 The rate computed from  Eq.~(\ref{eq:kzero}) is higher by 5 times than the experimental data \cite{Gibbons} or the calculation by Sturm \cite{Sturm}. This  is not surprising since the Umklapp process in metals are very different from those in dense plasmas where the ion lattice structure is absent.  
Lastly, we consider the partially degenerate plasma.  The damping is reduced in comparison to the degenerate case. This is natural since the ion-electron collisions are reduced in the partial degenerate case compared to the degenerate case \cite{sonprl, sonpla}.   This regime is the  most important in the application of the Raman compression in dense plasmas or the warm dense matter. For an example, consider a hydrogen plasma with the electron density of $10^{26} / \mathrm{cc}$ and the  temperature $400 \ \mathrm{eV}$. This case is shown in Fig.~\ref{fig:1}. The classical theory is not valid in this regime and it is necessary to use Eq.(\ref{eq:partial}).

The result obtained in this paper  could have many implications for processes involving the Langmuir wave in dense plasmas, warm dense matter and metals.  While the theory might need more refined adjustments such as the strong correlation, the local field correction, and the  exchange interaction, 
a complete profile of the damping rate is, if still rough, now available  as a function of the temperature, density and the wave vector $k$, which could be  important for various processes in dense plasmas~\cite{surface, photo, nano, wake, Fisch, Fisch2, sonprl,sonpre,Kue,IAW}.


Now, we discuss the implication of the result discussed above on the x-ray Raman compression. 
In dense plasmas, the higher the temperature is, the lower  the inverse bremsstrahlung is. 
 It is shown previously  that  the Landau damping of the plasmon generated from the pondermotive potential of a pump and a seed is greatly reduced in dense plasmas due to the electron quantum diffraction while the decay of the background noise plasmon could be very heavy due to a high electron temperature.  This suggests that the premature pump depletion from the background BRS is easy to suppress  while the BRS compression is still possible~\cite{sonlandau}. If  this is the case, the inverse bremsstrahlung and the FRS are the most important physics to check for the plausibility of the BRS compression. 
 The optimal parameter regime would be determined by this consideration.

For $T_e =0$,  $\gamma / \omega_{\mathrm{pe}}$ is shown, in this paper, to decrease with an increasing temperature, and for a fixed finite temperature, it is shown to decrease with an increasing density.    
If the FRS is too severe for $T_e=0$, the FRS could not be contained for $T_e > 0$ unless the density $n_e$ is higher. 
 Now, assume that the FRS is contained when $T_e=0 $ for a given $n_e$.  As the temperature increases, the FRS becomes stronger. Choose the maximum $T_e^{\mathrm{max}}$ such that the FRS is still tolerable. This  $T_e^{\mathrm{max}}$ might  be the optimal temperature since the inverse bremsstrahlung is minimal among the parameter regime where the FRS is contained.    

For a high electron temperature, the FRS might be  weak due to the enhanced Landau damping  instead of the damping from the Umklapp processes while the BRS is still strong due to the reduced Landau damping from the band gap \cite{sonlandau}.  In this case, the optimal temperature would be  the maximum temperature among the parameter regimes where the BRS plasmon is still a collective mode.  Whether the FRS is contained by the Umklapp process or the Landau damping will be important factor in the determination of the optimal physical parameter regime, which depends on the pulse duration and intensity.   We leave this question to a future researches.




 The ionic structure factor  depends strongly on the ion temperature in the warm dense matter as the ions experience  various phase transitions with the increasing temperature. As shown here, the plasmon damping is sensitive to the ionic structure factor. 
     This strong dependence  of the plasmon damping on the ionic structure factor  could be useful in the Raman compression, and it could also serve as a diagnostic in the X-ray Thomson scattering in the warm dense matter~\cite{Glenzer, Glenzer1}.  .  
In the case of the warm dense matter, the experiment to measure the damping rate  might be readily available using  the thin heated foil experiment \cite{Gibbons}. 
The effect of the phase transition on the plasmon decay is theoretically challenging, but very important for its practical application in the BRS x-ray compression and for other processes.  It would be interesting to see how the damping changes as the ionic structure factor changes.  

In addition to the theory proposed here,  the physical processes in dense plasmas could be very different from those of rare dense plasmas due to the electron degeneracy \cite{sonpla, sonprl}, the electron quantum  diffraction \cite{sonlandau}, and the band gap \cite{IAW, Kue}. The study and applications of those new physics could be exciting and interesting.

\bibliography{plasmon}

\end{document}